\documentclass[aps,prl,preprint,tightenlines,superscriptaddress,showpacs,byrevtex]{revtex4}
\usepackage{graphicx} 
\usepackage{dcolumn}  

\graphicspath{{ps}}

\def\be{\begin{equation}}
\def\ee{\end{equation}}

\newcommand{\lw}[1]{\smash{\lower1.7ex\hbox{#1}}}
\newcommand{\lww}[1]{\smash{\lower6.7ex\hbox{#1}}}
\newcommand{\bbbar}{\ensuremath{B\overline{B}} }
\newcommand{\btaunu}{\ensuremath{B^{-}\rightarrow\tau^{-}\overline{\nu}}}
\newcommand{\btag}{\ensuremath{B_{\rm tag}} }
\newcommand{\bsig}{\ensuremath{B_{\rm sig}} }

\newcommand{\eecl}{\ensuremath{E_{\text{ECL}}} }
\newcommand{\brvalue}{1.65}
\newcommand{\brstaterr}{^{+0.38}_{-0.37}}
\newcommand{\brsysterr}{^{+0.35}_{-0.37}}

\begin{document}


\preprint{\vbox{ \hbox{   }
                 \hbox{BELLE-CONF-0840}
}}

\title{ \quad\\[0.5cm]
  { \bf Measurement of $B^- \to \tau^- \overline{\nu}_{\tau}$ Decay With a Semileptonic Tagging Method\\}
  }


  
\affiliation{Budker Institute of Nuclear Physics, Novosibirsk}
\affiliation{Chiba University, Chiba}
\affiliation{University of Cincinnati, Cincinnati, Ohio 45221}
\affiliation{Department of Physics, Fu Jen Catholic University, Taipei}
\affiliation{Justus-Liebig-Universit\"at Gie\ss{}en, Gie\ss{}en}
\affiliation{The Graduate University for Advanced Studies, Hayama}
\affiliation{Gyeongsang National University, Chinju}
\affiliation{Hanyang University, Seoul}
\affiliation{University of Hawaii, Honolulu, Hawaii 96822}
\affiliation{High Energy Accelerator Research Organization (KEK), Tsukuba}
\affiliation{Hiroshima Institute of Technology, Hiroshima}
\affiliation{University of Illinois at Urbana-Champaign, Urbana, Illinois 61801}
\affiliation{Institute of High Energy Physics, Chinese Academy of Sciences, Beijing}
\affiliation{Institute of High Energy Physics, Vienna}
\affiliation{Institute of High Energy Physics, Protvino}
\affiliation{Institute for Theoretical and Experimental Physics, Moscow}
\affiliation{J. Stefan Institute, Ljubljana}
\affiliation{Kanagawa University, Yokohama}
\affiliation{Korea University, Seoul}
\affiliation{Kyoto University, Kyoto}
\affiliation{Kyungpook National University, Taegu}
\affiliation{\'Ecole Polytechnique F\'ed\'erale de Lausanne (EPFL), Lausanne}
\affiliation{Faculty of Mathematics and Physics, University of Ljubljana, Ljubljana}
\affiliation{University of Maribor, Maribor}
\affiliation{University of Melbourne, School of Physics, Victoria 3010}
\affiliation{Nagoya University, Nagoya}
\affiliation{Nara Women's University, Nara}
\affiliation{National Central University, Chung-li}
\affiliation{National United University, Miao Li}
\affiliation{Department of Physics, National Taiwan University, Taipei}
\affiliation{H. Niewodniczanski Institute of Nuclear Physics, Krakow}
\affiliation{Nippon Dental University, Niigata}
\affiliation{Niigata University, Niigata}
\affiliation{University of Nova Gorica, Nova Gorica}
\affiliation{Osaka City University, Osaka}
\affiliation{Osaka University, Osaka}
\affiliation{Panjab University, Chandigarh}
\affiliation{Peking University, Beijing}
\affiliation{Princeton University, Princeton, New Jersey 08544}
\affiliation{RIKEN BNL Research Center, Upton, New York 11973}
\affiliation{Saga University, Saga}
\affiliation{University of Science and Technology of China, Hefei}
\affiliation{Seoul National University, Seoul}
\affiliation{Shinshu University, Nagano}
\affiliation{Sungkyunkwan University, Suwon}
\affiliation{University of Sydney, Sydney, New South Wales}
\affiliation{Tata Institute of Fundamental Research, Mumbai}
\affiliation{Toho University, Funabashi}
\affiliation{Tohoku Gakuin University, Tagajo}
\affiliation{Tohoku University, Sendai}
\affiliation{Department of Physics, University of Tokyo, Tokyo}
\affiliation{Tokyo Institute of Technology, Tokyo}
\affiliation{Tokyo Metropolitan University, Tokyo}
\affiliation{Tokyo University of Agriculture and Technology, Tokyo}
\affiliation{Toyama National College of Maritime Technology, Toyama}
\affiliation{Virginia Polytechnic Institute and State University, Blacksburg, Virginia 24061}
\affiliation{Yonsei University, Seoul}
  \author{I.~Adachi}\affiliation{High Energy Accelerator Research Organization (KEK), Tsukuba} 
  \author{H.~Aihara}\affiliation{Department of Physics, University of Tokyo, Tokyo} 
  \author{D.~Anipko}\affiliation{Budker Institute of Nuclear Physics, Novosibirsk} 
  \author{K.~Arinstein}\affiliation{Budker Institute of Nuclear Physics, Novosibirsk} 
  \author{T.~Aso}\affiliation{Toyama National College of Maritime Technology, Toyama} 
  \author{V.~Aulchenko}\affiliation{Budker Institute of Nuclear Physics, Novosibirsk} 
  \author{T.~Aushev}\affiliation{\'Ecole Polytechnique F\'ed\'erale de Lausanne (EPFL), Lausanne}\affiliation{Institute for Theoretical and Experimental Physics, Moscow} 
  \author{T.~Aziz}\affiliation{Tata Institute of Fundamental Research, Mumbai} 
  \author{S.~Bahinipati}\affiliation{University of Cincinnati, Cincinnati, Ohio 45221} 
  \author{A.~M.~Bakich}\affiliation{University of Sydney, Sydney, New South Wales} 
% ...
\collaboration{The Belle Collaboration}

\noaffiliation

\begin{abstract}
We present a new measurement of the decay $B^{-}\rightarrow\tau^{-}\overline{\nu}_\tau$
with a semileptonic $B$ tagging method,
using a data sample containing $657\times 10^6$ \bbbar pairs collected at the $\Upsilon(4S)$
resonance with the Belle detector at the KEKB asymmetric $e^{+}e^{-}$ collider.
A sample of $\bbbar$ pairs are tagged by reconstructing one $B$ meson decaying
semileptonically. We detect the $B^-\to \tau^-\overline{\nu}_{\tau}$ candidate in
the recoil.
We obtain a signal with a significance of 3.8 standard deviations including systematics,
and measure the branching fraction to be
${\cal B}(B^{-}\rightarrow\tau^{-}\overline{\nu}_{\tau}) = 
 (\brvalue\brstaterr(\text{stat})\brsysterr(\text{syst})) \times 10^{-4}$.
This result confirms the evidence for $B^{-}\to\tau^-\overline{\nu}_\tau$ 
obtained in the previous Belle measurement with a hadronic $B$ tagging method.

\end{abstract}


\maketitle


{\renewcommand{\thefootnote}{\fnsymbol{footnote}}}
\setcounter{footnote}{0}

The purely leptonic decay $B^{-}\rightarrow\tau^{-}\overline{\nu}$~\cite{conjugate}
is of particular interest since it provides a direct measurement of the product of 
the Cabibbo-Kobayashi-Maskawa(CKM) matrix element $V_{ub}$~\cite{CKM} and
the $B$ meson decay constant $f_{B}$. In the Standard Model(SM), 
the branching fraction of the decay $B^{-}\rightarrow\tau^{-}\overline{\nu}$ 
is given by
\begin{equation}
 \label{eq:BR_B_taunu}
{\cal B}(B^{-}\rightarrow\tau^{-}\overline{\nu})_{SM} = \frac{G_{F}^{2}m_{B}m_{\tau}^{2}}{8\pi}\left(1-\frac{m_{\tau}^{2}}{m_{B}^{2}}\right)^{2}f_{B}^{2}|V_{ub}|^{2}\tau_{B},
\end{equation}
where $G_{F}$ is the Fermi coupling constant, $m_{\tau}$ and $m_{B}$ are
the $\tau$ lepton and $B$ meson masses, and $\tau_{B}$ is the $B^{-}$ lifetime.
The dependence on the lepton mass arises from helicity conservation, which 
suppresses the muon and electron channels.
Observation of $B^{-}\rightarrow\tau^{-}\overline{\nu}$ could provide 
a direct measurement of $f_B$.
Physics beyond the SM, such as supersymmetry or two-Higgs doublet models, could suppress
or enhance ${\cal B}(B^{-}\rightarrow\tau^{-}\overline{\nu})$ to levels several 
times as large as the SM expectation through 
the introduction of a charged Higgs boson \cite{Hou:1992sy,Baek:1999ch}.
The expected SM branching fraction from other experimental constraints is
$(0.93^{+0.11}_{-0.12})\times 10^{-4}$~\cite{CKMfitter2007}.

The previous Belle measurement\cite{ikado-2006-97} reported the first evidence of 
 $B^{-}\rightarrow\tau^{-}\overline{\nu}$ decay
 with a significance of $3.5$ standard deviations ($\sigma$),
 and measured the branching fraction to be
${\cal B}(B^{-}\rightarrow\tau^{-}\overline{\nu}_{\tau}) = 
 (1.79^{+0.56}_{-0.49}(\mbox{stat})^{+0.46}_{-0.51}(\mbox{syst})) \times 10^{-4}$, 
 using a full reconstruction tagging method.
The BaBar Collaboration has reported a search for $B^{-}\rightarrow\tau^{-}\overline{\nu}$
decay with hadronic tagging~\cite{Aubert:2007} and
semileptonic tagging~\cite{Aubert:2007_semil} using $383 \times 10^6$ \bbbar pairs. 
They report a 2.6 $\sigma$ excess, combining the two measurements.
No statistically significant enhancement relative to the SM expectation has been observed
in previous experimental studies.
To establish the $\btaunu$ signal, test consistency with the SM and search for a charged 
Higgs boson effect, we need more statistics.
In this paper, we present a new measurement of $B^{-}\rightarrow\tau^{-}\overline{\nu}_{\tau}$
from the Belle experiment with a semileptonic tagging method.

We use a $605~\textrm{fb}^{-1}$ data sample containing 
$657\times 10^{6}$ \bbbar pairs collected with the Belle detector
at the KEKB asymmetric energy $e^{+}e^{-}$ ($3.5$ on $8$ GeV) collider
operating at the $\Upsilon(4S)$ resonance ($\sqrt{s} = 10.58$ GeV)~\cite{KEKB}.
We also use a data sample of 68~fb$^{-1}$ taken with a center of mass energy
60~MeV below the nominal $\Upsilon(4S)$ (off-resonance) mass for a background study.
The Belle detector is a large-solid-angle magnetic
spectrometer that
consists of a silicon vertex detector (SVD),
a 50-layer central drift chamber (CDC), an array of
aerogel threshold Cherenkov counters (ACC), 
a barrel-like arrangement of time-of-flight
scintillation counters (TOF), and an electromagnetic calorimeter
(ECL) comprised of CsI(Tl) crystals located inside 
a superconducting solenoid coil that provides a 1.5~T
magnetic field.  An iron flux-return located outside of
the coil is instrumented to detect $K_L^0$ mesons and to identify
muons (KLM).  
Two inner detector configurations were used. A 2.0 cm beampipe
and a 3-layer silicon vertex detector was used for the first sample
of 152 $\times 10^6 B\bar{B}$ pairs, while a 1.5 cm beampipe, a 4-layer
silicon detector and a small-cell inner drift chamber were used to record  
the remaining 505 $\times 10^6 B\bar{B}$ pairs \cite{svd2}.  
The detector is described in detail elsewhere~\cite{Belle}.

We use a detailed Monte Carlo (MC) simulation based on GEANT~\cite{GEANT} 
to determine the signal selection efficiency and study the background.
In order to reproduce effects of beam background, data taken with random
triggers for each run period are overlaid on simulated events. 
The $B^{-}\rightarrow\tau^{-}\bar{\nu}_{\tau}$ signal decay is generated
by the EvtGen package~\cite{EvtGen}.
To model the background from $e^+e^- \to B\overline{B}$ and continuum 
$q\overline{q}~(q = u, d, s, c)$ production processes, large 
$B\overline{B}$ and $q\overline{q}$ MC samples 
corresponding to about four times the data sample are used.
We also use MC samples for various rare $B$ decay processes such as charmless 
hadronic, radiative, electroweak decays and $b \to u$ semileptonic decays.
We also use a MC sample for $e^+e^-\to \tau^+\tau^-$ events.

The strategy adopted for this analysis is same as in the previous measurements.
We reconstruct one of the $B$ mesons decaying semileptonically
(referred to hereafter as $\btag$)
and compare the properties of the remaining particle(s) in the event ($\bsig$) to those
expected for signal and background.
In order to avoid experimental bias, 
the signal region in data is not examined until the event 
selection criteria are finalized.

We reconstruct the \btag in $B^-\to D^{*0}\ell^-\overline{\nu}$ and
$B^-\to D^{0}\ell^-\overline{\nu}$ decays.
For $D^{*0}$ reconstruction, we use $D^{*0}\to D^{0}\pi^0$ and $D^{0}\gamma$ decays.
$D^0$ mesons are reconstructed in $K^-\pi^+$, $K^-\pi^+\pi^0$ and $K^-\pi^+\pi^-\pi^+$.
For $\bsig$, we use $\tau^-$ decays to only one charged particle and neutrinos: 
$\tau^- \to \ell^- \overline{\nu}_\ell \nu_\tau$ and
$\tau^- \to \pi^- \nu_\tau$.
We require that no charged particle or $\pi^0$ remain in the event after removing
the particles from the \btag and \bsig candidates.

Charged particles are selected from well measured tracks reconstructed with the CDC and SVD
originating from the interaction point.
Electron candidates are identified based on a likelihood calculated using 
$dE/dx$ in CDC, the response of ACC, ECL shower shape and the 
ratio of the ECL energy deposit and the track momentum.
Muon candidates are selected using KLM hits associated to a charged track.
Both muons and electrons are selected with efficiency greater than 90\% 
in the momentum region above 1.2 GeV/$c$, and 
misidentification rates of less than 0.2\%(1.5\%) for electrons (muons). 
After selecting leptons, we distinguish charged kaons from pions 
based on a kaon likelihood derived from the TOF, ACC, and $dE/dx$ measurements in the CDC.
The typical kaon identification efficiency is more than 85\% and 
the probability of misidentifying pions 
as kaons is about 8\%.
Photons are identified as isolated ECL clusters that are not matched to any charged track.
$\pi^0$ candidates are selected from pairs of photons with invariant mass between 0.118 and
0.150 GeV/c$^2$.
Photons from $\pi^0$ candidates used in $D^0$ meson reconstruction and photons from
$D^{*0} \to D^{0}\gamma$ decays are required to pass the following energy requirements:
50 MeV for the barrel, 100 MeV for the forward endcap and 150 MeV for the backward endcap.
For low momentum $\pi^0$'s from $D^{*0}\to D^{0}\pi^0$ decay,
we require the photon energy to be greater than 30 MeV.

$D^0$ meson candidates are selected from combinations of charged kaon, pion and $\pi^0$
candidates.
We require the invariant mass of $D^0$ candidates to be in the range
$1.851 < M_{D^0} < 1.879$~GeV/c$^2$ for $D^0\to K^-\pi^+$ and $K^-\pi^+\pi^-\pi^+$ decays,
and $1.829 < M_{D^0} < 1.901$~GeV/c$^2$ for $D^0\to K^-\pi^+\pi^0$ decay.
$D^{*0}$ candidates are selected by combining the $D^0$ candidates with low momentum $\pi^0$ 
candidates and photons. For $D^{*0}$ candidates, we require the mass difference
$M_D^{*0} - M_{D^0}$ to be in the range $0.1389 < M_D^{*0} - M_{D^0} < 0.1455$~GeV/c$^2$
and $0.123 < M_D^{*0} - M_{D^0} < 0.165$~GeV/c$^2$ for
$D^{*0}\to D^0\pi^0$ and $D^{*0}\to D^0\gamma$ decays, respectively.
The regions correspond to 3$\sigma$ from the nominal $D^0$ mass or from the nominal mass difference.
We select \btag candidates using the lepton momentum $P^*_\ell$ and the cosine of the angle between 
the direction of the \btag momentum and the direction of the momentum sum of the $D^{(*)0}$ 
and the lepton $\cos\theta_{B-D^{(*)}\ell}$. This angle is calculated using
    $\cos\theta_{B-D^{(*)}\ell} = (2E_{\text{beam}}E_{D^{(*)}\ell}-m^2_B-m^2_{D^{(*)}\ell})/(2 P_B\cdot P_{D^{(*)}\ell})$,
where $E_{D^{(*)}\ell}$, $P_{D^{(*)}\ell}$ and $M_{D^{(*)}\ell}$ are the energy sum, momentum sum 
and invariant mass of the $D^{(*)0}$ and lepton. All parameters are calculated in the cms.
Here $m_B$ is the nominal $B^-$ mass~\cite{pdg2006}, and the $P_B$ is the momentum of 
$B$ meson in the cms calculated with $P_B = \sqrt{E_{\text{beam}}^2 - m_B^2}$.
For the signal side track, we require the momentum $P^*_{\tau\to X}$ to be in the region consistent
with a $B \to \tau\overline{\nu}$ decay. 
The selection criteria for \btag and \bsig are optimized for each of the $\tau$ decay modes,
because the background levels and the background components are mode-dependent.
The optimization is done so that the figure of merit $s/\sqrt{s+n}$ is maximized,
where $s$ and $n$ are the number of signal and background events 
in the signal enhanced region with remaining energy in the ECL less than 0.2 GeV,
which will be described later,
calculated with the signal branching fraction of $1.79 \times 10^{-4}$.
For leptonic $\tau$ decays, the dominant background is \bbbar pair events
correctly tagged by a semileptonic decay.
Therefore loose selection criteria are chosen to maintain high signal efficiency:
$0.5 < P^*_{\ell} < 2.5$ GeV/c, $-2.1 < \cos\theta_{B-D^{*}\ell}<1.3$ for $D^{*}$ mode
or $-2.6 < \cos\theta_{B-D\ell}<1.2$ for $D^0$ mode, and $0.3 \text{ GeV/c} < P^*_{\tau\to X}$.
For hadronic $\tau$ decay modes, there is more background 
from $e^+e^- \to q\overline{q}$ continuum and combinatoric $D^{(*)0}\ell$ background.
Tighter criteria are used to reduce such backgrounds:
$1.0 < P^*_{\ell} < 2.2$ GeV/c, $-1.1 < \cos\theta_{B-D^{(*)0}\ell}<1.1$,
and $1.0 < P^*_{\tau\to X} < 2.4 $ GeV/c.
The upper bound on $P^*_{\tau\to X}$ is introduced to reject two body $B$ decays.
In addition, we suppress continuum background by requiring the cosine of the angle between the signal 
side pion track and the thrust axis of the \btag $\cos\theta_{ \text{thr} }$ to be less than 
0.9.

The most powerful variable for separating signal and background is the
remaining energy in the ECL, denoted $E_{\rm ECL}$, which is the sum of
the energies of ECL clusters that are not associated with 
particles from the \btag and \bsig candidates.
We require a minimum energy threshold of 50 MeV for the barrel and 100 
MeV for the forward, and 150 MeV for the backward endcap ECL.
A higher threshold is used for the endcap ECL because the effect of beam
background is more severe.
For signal events, $E_{\rm ECL}$ must be either zero or a small value 
arising from residual beam background hits, therefore, signal events peak at 
low $E_{\rm ECL}$.
On the other hand, background events are distributed toward higher 
$E_{\rm ECL}$ due to the contribution from additional particles.
We select candidate events in the range $\eecl < 1.2$ GeV for further analysis.

\begin{figure}
\begin{center}
        \includegraphics[width=0.5\textwidth]{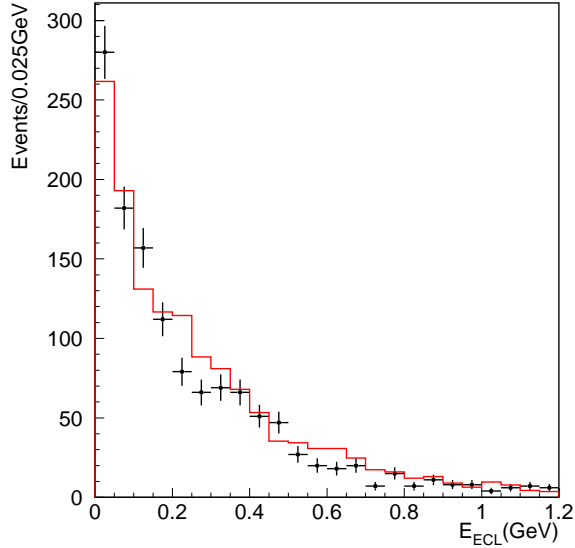}
    \caption{
      \eecl distribution for double semileptonic tagged events.
      The points with error bars are for data and the solid histogram is the MC expectation.
      }
    \label{fig:doubletag}
\end{center}
\end{figure}

The dominant background contributions are from \bbbar pair and continuum processes,
which are estimated to be 86\% (59\%), 11\% (36\%) for the leptonic (hadronic) $\tau$ decays 
from MC, respectively.
Background from rare $B$ decays and $\tau$ pair events are estimated from MC to be 
small (3\% for leptonic $\tau$ decays and 4\% for hadronic $\tau$ decay)
and fixed in the fit to the MC expectation.
In order to validate the $E_{\rm ECL}$ simulation, we use a control sample
of double tagged events, where the $B_{\rm tag}$ is reconstructed
in a semileptonic decay as described above and $B_{\rm sig}$ is reconstructed in the decay chain,
$B^{-} \rightarrow D^{*0}\ell^{-}\bar{\nu}$ ($D^{*0}\rightarrow D^{0}\pi^{0}$),
followed by $D^0 \to K^- \pi^+$. 
The dominant source affecting the $E_{\rm ECL}$ distribution in the control sample is 
beam background hits and is common to the signal.
Figure~\ref{fig:doubletag} shows the $E_{\rm ECL}$ distribution in the
control sample for data and the MC simulation scaled to the same luminosity.
The background in this control sample is negligibly small.
The agreement between data and MC demonstrates the validity of the $E_{\rm ECL}$ simulation
in the signal MC.
The $\eecl$ background shape is validated using the $E_{\rm ECL}$ sideband region
defined by $0.4 < E_{\rm ECL} < 1.2$ GeV and off-resonance data.
We confirm the background shapes obtained from MC agree with the $E_{\rm ECL}$ shape
of these background data samples.

\begin{figure}[bh]
    \begin{center}
	\includegraphics[width=0.8\textwidth]{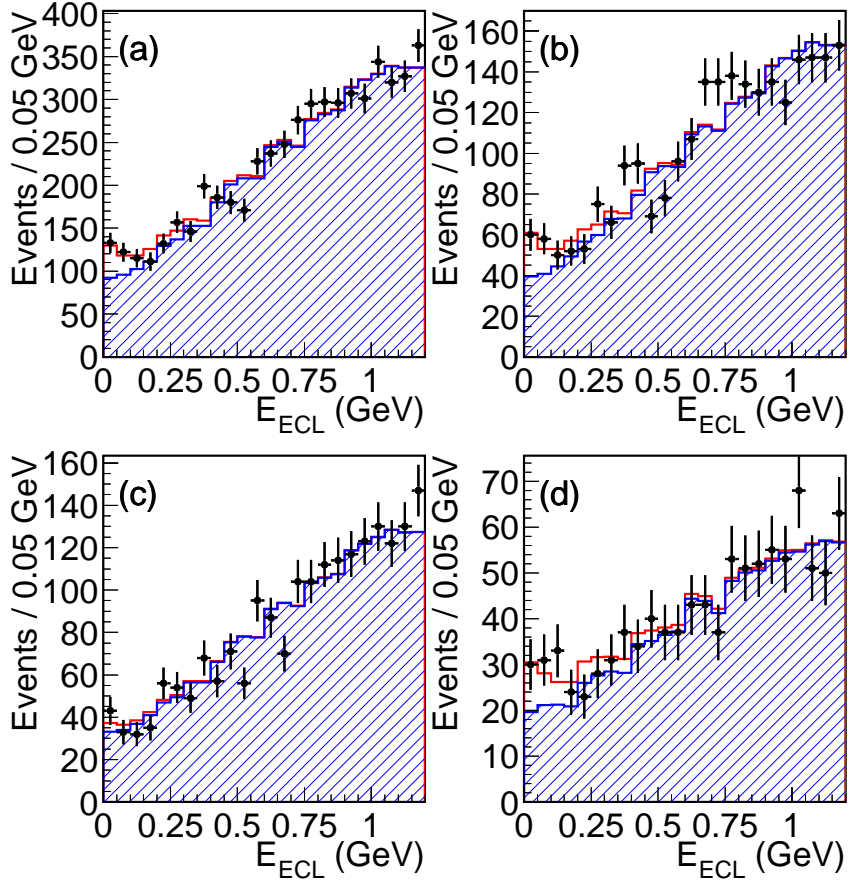}
	\caption{\eecl distribution of semileptonic tagged events 
	  with the fit result for 
	  (a) all $\tau$ decay modes combined,
	  (b) $\tau^- \to e^-\overline{\nu}_e\nu_\tau$,
	  (c) $\tau^- \to \mu^-\overline{\nu}_\mu\nu_\tau$ and
	  (d) $\tau^- \to \pi^-\nu_\tau$.
	  The points with error bars are data. The hatched histogram
	  and solid open histogram are the background and the signal, 
	  respectively.
	  }
	\label{fig:eecl_fit}
    \end{center}
\end{figure}

After finalizing the signal selection criteria, the signal region is examined.
The number of signal events is extracted from an extended maximum likelihood fit to the \eecl
distribution.
Probability density functions (PDFs) for each $\tau$ decay mode 
are constructed from the MC simulation.
We use \eecl histograms obtained from MC samples for each of the signal and
the background components. 
The PDFs are combined into a likelihood function,
\begin{equation}
{\cal L} = \frac{e^{-\sum_j n_j}}{N!}
\prod_{i=1}^{N}\sum_j n_j f_j(E_{i})
\end{equation}
where $E_{i}$ is the $E_{\rm ECL}$ value in the $i$th event, $N$ is the total number 
of events in the data, and $n_{j}$ is the yield of the $j$th component, where $j$ is an 
index for the signal and background contributions.
In the final fit, four parameters are floated: the signal yield
and the sum of \bbbar and continuum backgrounds for the three $\tau$ decay modes.
We combine $\tau$ decay modes by constraining the ratios of the signal 
yields to the ratio of reconstruction efficiencies obtained from MC.
Figure~\ref{fig:eecl_fit} shows the $E_{\rm ECL}$ distribution obtained with the fit results.
\eecl distributions for each $\tau$ decay mode is also shown.
We see a clear excess of signal events in the region near $E_{\rm ECL}\sim 0$.
We obtain the signal yield to be $n_{\rm s} = 154^{+36}_{-35}$ .
The branching fraction is calculated as
${\cal B} = n_{\rm s}/(2\cdot\varepsilon\cdot N_{B^{+}B^{-}})$,
where $\varepsilon$ is the reconstruction efficiency including the tagging efficiency
and the branching fraction of the $\tau$ decay modes and 
$N_{B^{+}B^{-}}$ is the number of $\Upsilon(4S)\rightarrow B^{+}B^{-}$ 
events, assuming $N_{B^{+}B^{-}} = N_{B^{0}\overline{B}^{0}}$.
Table~\ref{tab:fit_result} shows the signal yields and the branching fractions
obtained from separate fits for each $\tau$ decay mode and fits with all three modes combined.

\begin{table}
 \begin{center}
    \begin{tabular}{lcccc} \hline \hline
Decay Mode & Signal Yield   &  $\varepsilon$ & $\cal B$  \\ \hline
$\tau^-\to e^{-}\nu\bar{\nu}_{\tau}$       &$78^{+23}_{-22}$  & $5.9\times 10^{-4}$ & $(2.02^{+0.59}_{-0.56})\times 10^{-4}$ \\
$\tau^-\to \mu^{-}\nu\bar{\nu}_{\tau}$     &$15^{+18}_{-17}$  & $3.7\times 10^{-4}$ & $(0.62^{+0.76}_{-0.71})\times 10^{-4}$ \\
$\tau^-\to\pi^{-}\nu_{\tau}$               &$58^{+21}_{-20}$  & $4.7\times 10^{-4}$ & $(1.88^{+0.70}_{-0.66})\times 10^{-4}$ \\
\hline
Combined                                   &$154^{+36}_{-35}$ & $14.3\times 10^{-4}$ & $(1.65^{+0.38}_{-0.37})\times 10^{-4}$ \\
\hline\hline
    \end{tabular}
    \caption{
      Results of the fit for signal yields and branching fractions.
      }
   \label{tab:fit_result}
 \end{center}
\end{table}

Systematic errors for the measured branching fraction are associated with 
the uncertainties in the signal yield, efficiencies and number of $B^{+}B^{-}$ pairs. 
The systematic errors for the signal yield arise from the uncertainties in the PDF shapes
for the signal ($^{+3.1}_{-3.2}$\%) and for the background ($^{+11.8}_{-11.2}$\%)
which are dominated by MC statistics. 
For the latter, uncertainties in the branching fractions of $B$ decay modes that 
peak at $\eecl=0$ such as $B^-\to D^0\ell^+\nu$ with $D^0\to K_L^0 \pi^0, K_L^0 K_L^0$
and so on ($^{+4.2}_{-8.4}$)\%, 
as well as uncertainties in the background from rare $B$ decays and
$\tau$ pair events ($3.8$\%) are also taken into account.
We take a 11.6\% error as the systematic error associated with the tag reconstruction 
efficiency from the difference of yields between data and MC for the control sample.
This value includes the error in the branching fraction
${\cal B}(B^{-}\rightarrow D^{*0}\ell^{-}\bar{\nu})$, which we estimate from 
${\cal B}(B^{0}\rightarrow D^{*-}\ell^{+}\nu)$ in Ref.~\cite{pdg2006}
and isospin symmetry.
The systematic error in the signal efficiencies arises from the uncertainty in tracking
efficiency (1.0\%), particle identification efficiency (1.3\%),
branching fractions of $\tau$ decays (0.4\%), and MC statistics (0.9\%).
The systematic error due to the uncertainty in $N_{B^{+}B^{-}}$ is 1.4\%.
The total fractional systematic uncertainty is $^{+21}_{-22}\%$, 
and the branching fraction is
\begin{equation}
{\cal B}(B^{-}\rightarrow\tau^{-}\bar{\nu}_{\tau}) = (\brvalue\brstaterr(\text{stat})\brsysterr(\text{syst}))\times 10^{-4}.
\end{equation}
The significance of the observed signal is evaluated by 
$\Sigma = \sqrt{-2\ln({\cal L}_{0}/{\cal L}_{\rm max})}$ 
where ${\cal L}_{\rm max}$ and ${\cal L}_{0}$ denote the maximum likelihood 
value and likelihood value obtained assuming zero signal events, respectively.
The systematic uncertainty is convolved in the likelihood with a Gaussian distribution 
with a width corresponding to the systematic error of the signal yield.
We determine the significance of the signal yield to be 3.8.

In summary, we have measured
the decay $B^{-}\rightarrow\tau^{-}\overline{\nu}$ with \bbbar pair events tagged by 
semileptonic $B$ decays
using a data sample containing $657\times 10^6$ \bbbar pairs collected at the $\Upsilon(4S)$
resonance with the Belle detector at the KEKB asymmetric $e^{+}e^{-}$ collider. 
We measure the branching fraction to be 
$(\brvalue\brstaterr(\text{stat})\brsysterr(\text{syst}))\times 10^{-4}$.
with a significance of 3.8 standard deviations including systematics.
We confirm the evidence reported in the previous Belle measurement 
with \bbbar pair events tagged by hadronic $B$ decays.
Using the measured branching fraction and known values of $G_F$, $m_B$, 
$m_{\tau}$ and $\tau_B$~\cite{pdg2006}, the product
of the $B$ meson decay constant $f_B$ and the magnitude of the 
Cabibbo-Kobayashi-Maskawa matrix element $|V_{ub}|$ 
is determined to be 
$f_B |V_{ub}|= (9.7\pm1.1^{+1.0}_{-1.1}) \times 10^{-4}$ GeV.
The measured branching fraction is consistent with the SM expectation 
from other experimental constraints~\cite{CKMfitter2007}.

We thank the KEKB group for the excellent operation of the
accelerator, the KEK cryogenics group for the efficient
operation of the solenoid, and the KEK computer group and
the National Institute of Informatics for valuable computing
and SINET3 network support. We acknowledge support from
the Ministry of Education, Culture, Sports, Science, and
Technology of Japan and the Japan Society for the Promotion
of Science; the Australian Research Council and the
Australian Department of Education, Science and Training;
the National Natural Science Foundation of China under
contract No.~10575109 and 10775142; the Department of
Science and Technology of India; 
the BK21 program of the Ministry of Education of Korea, 
the CHEP SRC program and Basic Research program 
(grant No.~R01-2005-000-10089-0) of the Korea Science and
Engineering Foundation, and the Pure Basic Research Group 
program of the Korea Research Foundation; 
the Polish State Committee for Scientific Research; 
the Ministry of Education and Science of the Russian
Federation and the Russian Federal Agency for Atomic Energy;
the Slovenian Research Agency;  the Swiss
National Science Foundation; the National Science Council
and the Ministry of Education of Taiwan; and the U.S.\
Department of Energy.


\begin{thebibliography}{99}
\bibitem{conjugate}
Throughout this paper, 
the inclusion of the charge conjugate mode decay is implied
unless otherwise stated.
\end{thebibliography}
\end{document}